\documentclass[aps,prd,twocolumn,superscriptaddress,groupedaddress,nofootinbib]{revtex4}

\usepackage{amsmath,amssymb,bm,color,dcolumn,mathrsfs,tabularx,graphicx}
\usepackage[hidelinks=true]{hyperref}
\hypersetup{colorlinks=true,linkcolor=blue,citecolor=blue,urlcolor=blue}
\usepackage{Macro} 

\def\T{\Theta}
\def\estt{\widehat{\tau}}
\def\hA{\widehat{A}}
\def\SMICA{{\tt SMICA}}
\def\NILC{{\tt NILC}}

\begin{document}

\title{Constraints on Patchy Reionization from Planck CMB Temperature Trispectrum}

\author{Toshiya Namikawa}
\affiliation{Department of Physics, Stanford University, Stanford, California 94305, USA}
\affiliation{Kavli Institute for Particle Astrophysics and Cosmology, SLAC National Accelerator Laboratory, 2575 Sand Hill Road, Menlo Park, California 94025, USA}

\date{\today}

\begin{abstract}
We present constraints on the patchy reionization by measuring the trispectrum of 
the {\it Planck} 2015 cosmic microwave background (CMB) temperature anisotropies. 
The patchy reionization leads to anisotropies in the CMB optical depth, and 
the statistics of the observed CMB anisotropies is altered. 
We estimate the trispectrum of the CMB temperature anisotropies to constrain spatial variation of the optical depth.
We show that the measured trispectrum is consistent with that from the standard lensed CMB simulation at $2\sigma$.
While no evidence of the patchy reionization is found in the Planck 2015 temperature trispectrum, 
the CMB constraint on the patchy reionization is significantly improved from previous works. 
Assuming the analytic bubble-halo model of Wang and Hu (2006), the constraint obtained in this work rules out
the typical bubble size at the ionization fraction of $\sim0.5$ as $R\agt 10$ Mpc.
Further, our constraint implies that large-scale $B$-modes from the patchy reionization are not a significant 
contamination in detecting the primordial gravitational waves of $r\agt0.001$ 
if the $B$ mode induced by the patchy reionization is described by Dvorkin et al. (2009).
The CMB trispectrum data starts to provide meaningful constraints on the patchy reionization. 
\end{abstract}

\maketitle


\section{Introduction} \label{intro}

The cosmic microwave background (CMB) has been measured very precisely and provided valuable implications for cosmology. 
The CMB anisotropies have been also used to detect the secondary effects such as gravitational weak lensing 
(e.g. Refs.~\cite{BKVIII,ACT16:phi,P15:phi,PB14:phi,Story:2014hni}) 
and thermal/kinetic Sunyaev-Zel'dovich (tSZ/kSZ) effects (e.g. Refs.~\cite{George:2014oba,ACTkSZ:2017}). 
In the near future, multiple CMB experiments will significantly improve the sensitivity to 
temperature and polarization anisotropies and to the secondary effects on CMB. 

Another important secondary effect which is not yet detected by CMB measurements is the anisotropies
of the CMB optical depth induced by the patchy reionization. 
The reionization is an inhomogeneous process and the directional dependence of the optical depth 
changes the statistics of the observed CMB fluctuations. 

The constraint on the reionization is limited and the details of the reionization process are still unknown. 
The Gunn-Peterson trough in quasar spectra indicates that the transition to 
a fully ionized universe occurred around $z_{\rm re}\sim 6$ (e.g. Ref.~\cite{Fan:2006dp}). 
The 21 cm measurements by EDGES put a lower bound on the duration of the reionization as 
$\Delta z\agt 0.4$ \cite{Bowman:2012,Monsalve:2017mli}.
The constraint on the kSZ power obtained from the South Pole Telescope suggests $\Delta z\alt 5.4$ \cite{George:2014oba}.
The CMB temperature/polarization power spectra observed by Planck provide the best constraints to date 
on the mean optical depth, $\tau=0.058\pm0.012$ \cite{P16:lowl}.  
The Planck CMB angular power spectra also suggest that the reionization occurs at $z_{\rm re}\sim 8$ and 
the duration of the reionization is at most $\Delta z\sim 2.8$ \cite{P16:reion}. 
However, the actual reionization is expected to be highly inhomogeneous whose details are still unknown. 
The constrains on the inhomogeneity of the reionization, and thus anisotropies of the optical depth, 
will provide valuable implications for the reionization process. 

In this paper, we measure the trispectrum of the CMB temperature anisotropies induced by the anisotropies 
of the optical depth, $\tau(\hatn)$. We apply the reconstruction method proposed by Ref.~\cite{Dvorkin:2008tf}. 
The previous effort by Ref.~\cite{Gluscevic:2012qv} used 
the temperature and polarization data from WMAP and constrained the trispectrum by the patchy reionization. 
The Planck data can address the temperature fluctuations at a smaller scale than the WMAP data, 
and significantly improves the constraints on the trispectrum. 

This paper is organized as follows. 
\sec{data} summarizes data used in the paper and \sec{analysis} explains 
our method to measure the trispectrum induced by the patchy reionization. 
\sec{results} shows our main results including implications for patchy reionization. 
\sec{discuss} discuss possible systematics in the measurement of the trispectrum. 
\sec{summary} is devoted to summary of our results. 
Verification of our pipeline is given in the Appendix. 

\section{Data and Simulation} \label{data}

We use public Planck 2015 temperature data stored at NERSC. 
\footnote{http://crd.lbl.gov/departments/computational-science/c3/c3-research/cosmic-microwave-background/cmb-data-at-nersc/}
\footnote{The Planck 2015 data also contains the CMB polarization maps, but the polarization data is 
too noisy to cross-check the results using the temperature data. 
Thus, we only use the temperature data to reconstruct the patchy reionization signals.}
In baseline analysis, we use the \SMICA\ temperature map provided by Planck collaboration, 
which is obtained by cleaning the foregrounds using multiple frequency maps. 
The temperature map is provided on the sphere with $\text{{\tt NSIDE}}=2048$. 
We also employ the corresponding FFP8.1 Monte Carlo (MC) simulations of 
the lensed CMB signals and noises \cite{FFP8}, and the effective beam function.

We employ the Planck lensing mask used for measuring gravitational lensing \cite{P15:phi} because 
the measurement of the optical depth anisotropies is similar to that of lensing. 
This mask is an extension of the standard \SMICA\ mask and additionally masks Galactic CO emission and 
residual point sources as they would further contribute to the CMB trispectrum. 
The lensing mask is apodized with $\theta=1$ deg using {\tt process\_mask} provided by 
the {\tt Healpix} package \cite{Gorski:2004by}.

To test the residual foreground components, the Planck collaboration also provides 
alternative foreground-cleaned maps but assuming Planck 2013 cosmology (FFP8) for the CMB signal components. 
When we use the FFP8 simulation, the MC signal components are multiplied by $1.0134$ 
as suggested by the Planck team to match the simulation with observation at the power spectrum level. 

As pointed out in Ref.~\cite{P15:phi}, the noise components are slightly underestimated in the simulation. 
To estimate the noise level of the \SMICA\ temperature map, we use the half-mission (MH) half-differenced map
which is expected to be dominated by noise as shown in \cite{P15:compsep:cmb}. 
\footnote{Other split data are available, but the corresponding simulations are not published in NERSC.}
We compute the CMB temperature power spectrum from the HM half-differenced map 
and compare it with the corresponding simulation, 
finding that the simulated noise is roughly $3$\% smaller than 
the noise estimated from the HM half-differenced map. 
Thus, in our analysis, we scale the \SMICA\ simulated noise map by $1.03$. 
This noise assumption is tested latter in \sec{discuss}.

The residual foreground components are not provided by the Planck collaboration. 
Similar to \cite{P15:phi}, we fill the missing components by adding a random Gaussian field 
whose power corresponds to the difference between the observed and simulated (signal+noise) spectra. 
We test the residual foreground in \sec{discuss}.

\section{Method} \label{analysis}

In this section, we summarize our method to estimate the power spectrum of the optical depth. 
We verify our method in the Appendix \ref{verify}. 

\subsection{Effect of optical depth anisotropies on CMB temperature}

Free electrons scatter CMB photons and observed small-scale temperature fluctuations, $\hT$, are 
suppressed by $\E^{-\tau(\hatn)}$ where $\tau(\hatn)$ is the line-of-sight 
optical depth on CMB. Denoting the temperature fluctuations at the recombination as $\T$, 
the observed temperature anisotropies become
\al{
	\hT(\hatn) &= \E^{-\tau(\hatn)}\T(\hatn) 
		\notag \\
		&= \T(\hatn) - \tau(\hatn)\T(\hatn) + \mC{O}(\tau^2) \,.
}
The second term leads to the mode coupling between the temperature anisotropies in the harmonic space. 
The estimator to reconstruct the optical depth anisotropies is then given as a convolution of 
the temperature fluctuations as similar to the lensing reconstruction. 

\subsection{Quadratic estimator for the inhomogeneous optical depth}

The quadratic estimator for the inhomogeneous optical depth has been derived in Refs.~ 
\cite{Dvorkin:2009ah,Gluscevic:2012qv}. Here we briefly summarize the estimator. 
We first compute the convolution of the temperature fluctuations. 
We apply the foreground mask to the temperature map, and
transform the masked temperature map to the harmonic coefficients, $\T_{\l m}$.
The multipole coefficients are then beam-deconvolved and 
multiplied by a filtering function to optimize the estimator. 
The filtered multipoles are transformed into angular space, 
and are used to form an unnormalized quadratic estimator; 
\al{
	\bar{\tau}(\hatn) = 
		\left[\sum_{\l m} \frac{Y_{\l m}(\hatn)\CTT_\l\hT_{\l m}}{\CTT_\l+N^{\T\T}_\l}\right]
		\left[\sum_{\l m} \frac{Y_{\l m}(\hatn)\hT_{\l m}}{\CTT_\l+N^{\T\T}_\l}\right]
	\,. \label{Eq:btau}
}
Here $Y_{\l m}(\hatn)$ is the spin-zero spherical harmonics, 
$\CTT_\l$ is the lensed temperature spectrum and $N^{\T\T}_\l$ is the sum of noise and residual components. 

The ensemble average of the above estimator, the mean-field bias, is nonzero if there are e.g. 
Galactic and point source masks, and anisotropic instrumental noise. 
We compute the mean-field bias, $\ave{\bar{\tau}}$, by averaging over the simulation realizations. 
We then subtract it from the above estimator, and obtain an unbiased estimator of $\tau_{LM}$;
\al{
	\estt_{LM} = A_L(\bar{\tau}_{LM}-\ave{\bar{\tau}_{LM}}) 
	\,, \label{Eq:estt}
}
where the estimator normalization $A_L$ is given as
\al{
	A_L^{-1} &= \sum_{\l\l'} \frac{(\CTT_\l)^2}{(\CTT_\l+N^{\T\T}_\l)(\CTT_{\l'}+N^{\T\T}_{\l'})}
		\notag \\
		&\times \frac{(2\l+1)(2\l'+1)}{16\pi}\Wjm{L}{\l}{\l'}{0}{0}{0}^2
	\,.
}

\subsection{Estimating power spectrum of optical depth}

The estimator of the optical depth is given as a quadratic in CMB anisotropies. 
As similar to the gravitational lensing analysis, the power spectrum of the estimator $\estt$ 
is given as the four-point correlation of CMB anisotropies and can be broken into 
disconnected and connected (trispectrum) parts as
\al{
	\ave{|\estt_{LM}|^2} = \ave{|\estt_{LM}|^2}_{\rm d} + \ave{|\estt_{LM}|^2}_{\rm c} 
	\,. \label{Eq:4pt}
}
The most significant contribution comes from the disconnected part, usually referred to as 
the disconnected bias or Gaussian bias. This part is given by 
\al{
	\ave{|\estt_{LM}|^2}_{\rm d} &= 
		\frac{1}{2}\ave{|\estt^{\T_{\bm 1}\T_{\bm 2}}_{LM}+\estt^{\T_{\bm 2}\T_{\bm 1}}_{LM}|^2}_{\bm 1,2} 
	\notag \\
	&\equiv N_L^{\rm d}
	\,. \label{Eq:4pt:d}
}
Here $\estt^{\T_{\bm 1}\T_{\bm 2}}_{LM}$ is the estimator where the temperature fluctuations, 
$\T_{\bm 1}$ and $\T_{\bm 2}$, are used in the first and second brackets in \eq{Eq:btau}, respectively. 
The index $i={\bm 1}, {\bm 2}$ denotes one of two sets of MC realizations 
and $\ave{\cdots}_{\bm i}$ is the ensemble average over the $i$th set of the MC realizations. 
The connected part contains the trispectrum induced by the patchy reionization. 
In addition, the lensing generates part of the connected part. 
From \eq{Eq:4pt}, the power spectrum estimator for the optical depth is given as
\al{
	\hC^{\tau\tau}_L = \frac{1}{2L+1}\sum_{M=-L}^L|\estt_{LM}|^2 - N_L^{\rm d} - N_L^{\rm lens} - N_L^{\rm mc}
	\,. \label{Eq:hCtt}
}
The third term is the contribution from the lensing. The last term is the correction for 
the MC noise due to the finite realizations of the simulation which is given as $N_L^{\rm d}$ 
divided by the number of realizations used to estimate the mean-field bias (see e.g. \cite{BenoitLevy:2013bc}). 

Since the disconnected bias has a large contribution, a mismatch between the observed and simulated 
temperature fluctuations leads to a significant bias in the power spectrum estimate. 
To accurately estimate the disconnected bias, we use a realization-dependent algorithm as similar 
to the lensing analysis. In this method, the disconnected bias is estimated by combining the observed 
and simulated temperature:
\al{
	\hN_L^{\rm d} 
		&\equiv \ave{|\estt_{LM}^{\T,\T_{\bm 1}}+\estt_{LM}^{\T_{\bm 1},\T}|^2}_1 
	\notag \\
		&- \frac{1}{2}\ave{|\estt^{\T_{\bm 1},\T_{\bm 2}}_{LM}+\estt^{\T_{\bm 2},\T_{\bm 1}}_{LM}|^2}_{\bm 1,2} 
	\,. \label{Eq:hN}
}
The above estimator is less sensitive to mismatches between the observed and simulated CMB fluctuations 
compared to the use of \eq{Eq:4pt:d} \cite{Namikawa:2012}. For example, if the simulated temperature 
has an error in the overall amplitude $(1+\delta)\T_{LM}$, the above estimator contains terms from 
$\mC{O}(\delta^2)$ while \eq{Eq:4pt:d} has contributions from $\mC{O}(\delta)$. 
In addition, the above estimator is useful to suppress spurious 
off-diagonal elements in the covariance matrix of $\hC_L^{\tau\tau}$ and to decrease the 
statistical uncertainties (e.g., Ref.~\cite{Hanson:2010rp}). 
The above estimator is derived as the optimal trispectrum estimator analogues to the lensing case 
\cite{Namikawa:2012}. 

The contribution from the lensing is also not negligible in estimating $\hC_L^{\tau\tau}$ \cite{Su:2011ff}. 
In the standard lensed CMB simulation (homogeneous reionization), 
the power spectrum of the quadratic estimator after 
subtracting the disconnected bias and MC noise is equivalent to the lensing contribution, $N_L^{\rm lens}$. 
Thus, we estimate the lensing contribution using the Planck standard simulation by subtracting 
$\hN^{\rm d}_L$ and $N^{\rm mc}_L$ from the power spectrum of $\estt$. 

\section{Results} \label{results}

\begin{figure}[t]
\bc
\includegraphics[width=8.9cm,height=6.5cm,clip]{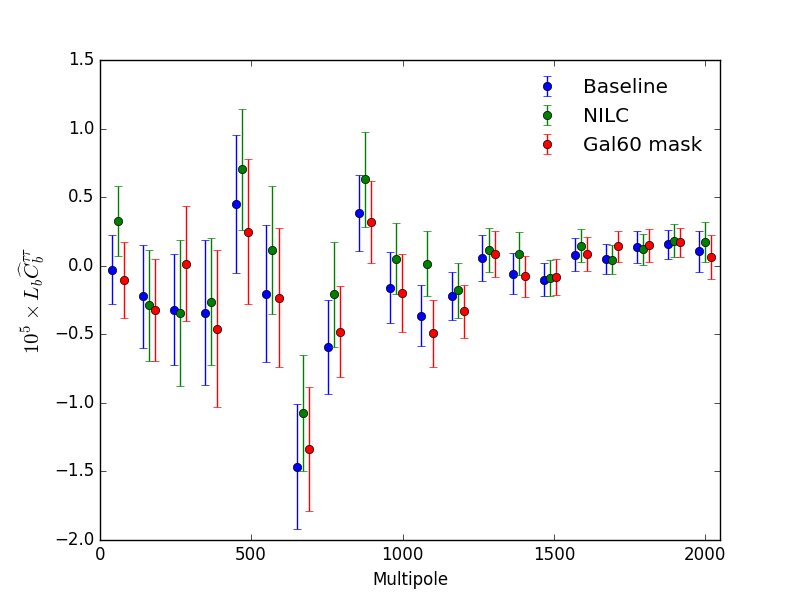} 
\caption{
The power spectrum of the optical depth reconstructed from the \SMICA\ Planck 2015 temperature map. 
We also show the case using a different foreground-cleaned temperature map and mask. 
For an illustrative purpose, the power spectra are multiplied by $10^5\times L$. 
}
\label{fig:cltt}
\ec
\end{figure}

Fig.~\ref{fig:cltt} shows the measurement of the angular power spectrum of the optical depth, 
$\hC_L^{\tau\tau}$, using the \SMICA\ Planck 2015 temperature map. 
We also compare the baseline result with the case using {\tt NILC} and with a different mask (Gal60). 
Note that there are correlations between the reconstructed power spectrum at different multipole bins and 
we should take into account the off-diagonal elements of the covariance to discuss the statistical 
significance of deviation from the null hypothesis.

To test whether the reconstructed power spectrum is consistent with the standard lensed CMB simulation, 
we estimate the probability-to-exceed (PTE) the value of $\chi$ and $\chi^2$.
The $\chi$ and $\chi^2$ values are estimated from (e.g. \cite{B2I})
\al{
	\chi   &= \sum_{b} \frac{\hC_b-C^{\rm fid}_b}{\sigma_b}
	\,, \notag \\
	\chi^2 &= \sum_{bb'} (\hC_b-C^{\rm fid}_b) \bR{Cov}^{-1}_{bb'}(\hC_{b'}-C^{\rm fid}_{b'}) 
	\,, \label{Eq:chi}
}
where $\sigma_b$ is the $1\sigma$ statistical error of the power spectrum, and
$\bR{Cov}_{bb'}$ is the power spectrum covariance estimated from the standard lensed CMB simulation. 
The ``measured'' spectrum $\hC_b$ is obtained from the real and simulation data. 
The fiducial spectrum $C_b^{\rm fid}$ is the simulation mean. 

We find that the $\chi$- and $\chi^2$-PTEs become $0.81$ and $0.15$, respectively. 
The reconstructed power spectrum is consistent with null hypothesis within $2\sigma$. 

Next, to show how significantly the Planck data improves the constraints on the patchy reionization 
from the previous works, 
we constrain a simple model where the power spectrum of the optical depth is 
given by \cite{Gluscevic:2012qv}
\al{
	C_L^{\tau\tau} = \left(\frac{A^\tau}{10^4}\right) \frac{4\pi}{L^2_c} \E^{-(L/L_c)^2} 
	\,. \label{Eq:model}
}
We derive constraint on the amplitude, $A^\tau$, at each correlation length as follows.
We first estimate the amplitude of the reconstructed power spectrum relative to the fiducial 
signal power spectrum ($A^\tau=1$) at each multipole bin.
The signal power spectrum is given as the sum of the input optical-depth power spectrum and N1 bias 
as similar to the lensing reconstruction. 
\footnote{
We find that the secondary contraction of the temperature trispectrum, 
usually referred to as the N1 bias \cite{Kesden:2003cc,Namikawa:2016d}, 
is significant and increases the total signal-to-noise of the optical depth power spectrum 
(see the Appendix \ref{verify}).
This additional contribution increases the total signal-to-noise and 
should be included to derive constraint on the optical depth anisotropies and 
to forecast expected constraints. 
}
Because both the input and N1 bias are proportional to $A^\tau$, we estimate the amplitude as
\al{
	\hA^\tau_b = \frac{\hC_b^{\tau\tau}}{(C_b^{\tau\tau}+N_b^{(1),\tau})|_{A^\tau=1}}
	\,.
}
Then we fit the measured $\hA^\tau$ to $A^\tau$ with the following likelihood:
\al{
	-2\ln\mC{L}(A^\tau) &= \ln\det(\mC{C}(A^\tau)) 
		\notag \\
		&+ \sum_{bb'} (\hA^\tau_b-A^\tau)\mC{C}^{-1}_{bb'}(A^\tau) (\hA^\tau_{b'}-A^\tau)
	\,.
}
where $\mC{C}_{bb'}$ is the covariance of $\hA^\tau_b$. 
We evaluate the covariance from the simulation. 
To take into account the cosmic variance from the inhomogeneous reionization, 
we correct the covariance by adding a diagonal components, $\mC{C}_{bb}=C_b^{\tau\tau}+N_b^{(1),\tau}$. 
The N1 bias, $N_b^{(1),\tau}$, is obtained from simulation where the effect of the patchy reionization is included. 
From the reconstructed power spectrum, we obtain the constraints on the amplitude of the optical-depth 
power spectrum at $2\sigma$ confidence level.

\begin{figure}[t]
\bc
\includegraphics[width=8.9cm,height=6.5cm,clip]{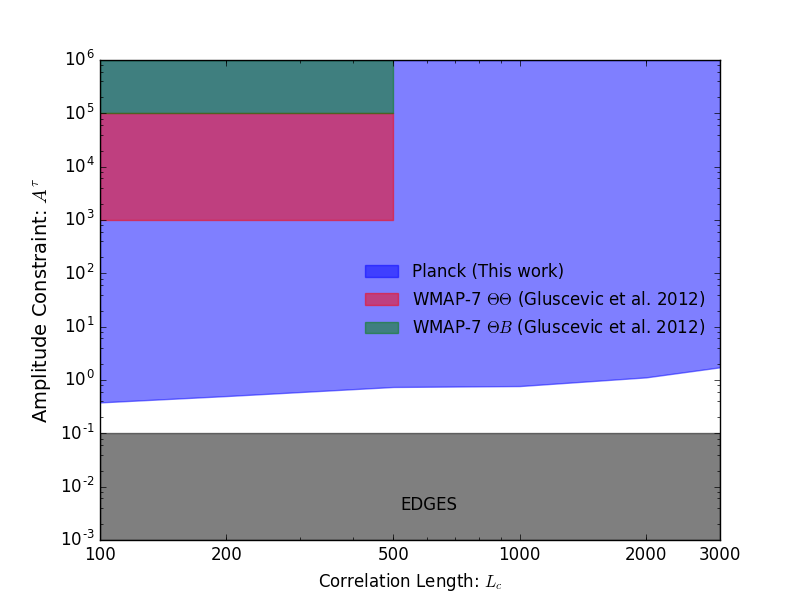} 
\caption{
Constraints on the amplitude of the optical-depth power spectrum of \eq{Eq:model} using 
the CMB trispectrum obtained in this work and in the previous work \cite{Gluscevic:2012qv}. 
The colored regions are excluded from observations.
The grey shaded region is derived from the lower limit of the reionization duration, 
$\Delta z\agt0.4$, by Ref.~\cite{Monsalve:2017mli} using EDGES. 
}
\label{fig:2dconst}
\ec
\end{figure}

Fig.~\ref{fig:2dconst} shows the results of the $2\sigma$ constraints on $A^\tau$ as a function of 
the correlation length $L_c$. 
We also show the constraints $A^\tau<10^3$ and $A^\tau<10^5$ ($L_c\leq 500$) obtained by 
Ref.~\cite{Gluscevic:2012qv} using the $\T\T\T\T$ and $\T B\T B$ trispectra, respectively.
Since the Planck data is sensitive to small angular scales, the Planck data significantly 
narrows the allowed region of the simple inhomogeneous reionization model. 
Following Ref.~\cite{Gluscevic:2012qv}, we translate the lower limit of the reionization duration 
by EDGES as $A^\tau/10^4=(\delta\tau)^2\sim (\Delta z\tau/z_{\rm re})^2\sim 0.1$ 
and show the corresponding constraint as a black shaded region. 
Note that the relationship between the duration of the reionization and the amplitude of 
the optical-depth power spectrum on a given scale is highly model dependent, and the lower bound 
denoted by ``EDGES'' should be used as an order of magnitude estimate. 

\begin{figure}[t]
\bc
\includegraphics[width=8.9cm,height=6.5cm,clip]{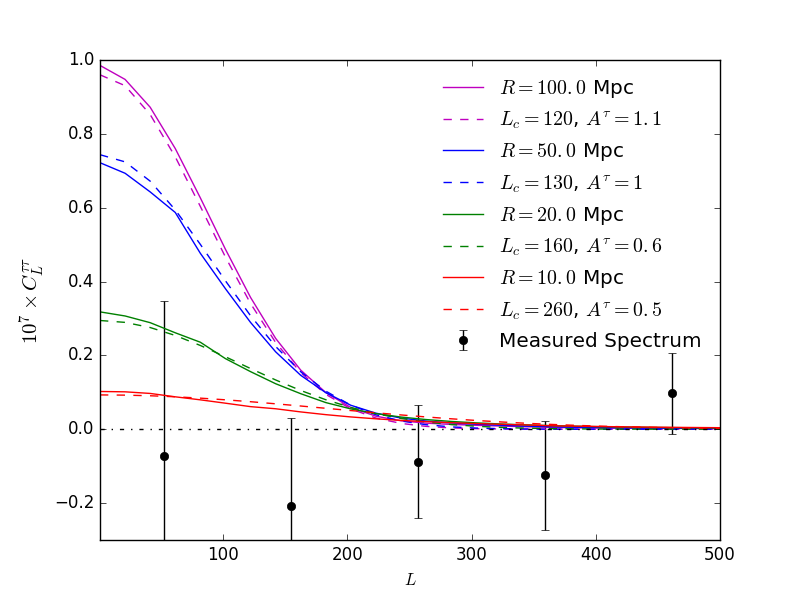} 
\caption{
Angular power spectra of the optical depth anisotropies predicted by the bubble halo model (solid lines). 
We also compare these spectra with the simple parametrization of \eq{Eq:model} with varying $A^\tau$ and $L_c$. 
}
\label{fig:anacltt}
\ec
\end{figure}

The above constraints on $A^\tau$ can be translated into 
that on more specific and physically-motivated models of the optical-depth anisotropies. 
Ref.~\cite{Wang:2006} (see also Ref.~\cite{Dvorkin:2008tf}) shows an analytic prediction of the $\tau$ power spectrum
based on the bubble halo model.
In their model, contributions to the power spectrum are decomposed into one and two bubble terms. 
The power spectrum depends on parameters of e.g. 
the evolution of the ionization fraction, $x_e(z)$, and 
the typical bubble size at $x_e\sim0.5$, $R$. 

Fig.~\ref{fig:anacltt} shows the angular power spectra of the optical depth anisotropies 
from the analytic bubble halo model. 
To compute the power spectra, we choose values of the cosmological parameters which are consistent with 
the Planck 2015 best fit results \cite{P15:main}. 
We employ the redshift-asymmetric power-law model of the evolution for the ionization fraction, $x_e(z)$ 
(see e.g. \cite{P16:reion})
with values of the model parameters consistent with the recent constraints \cite{P16:reion}. 
If $R\sim 10$ Mpc, their power spectrum corresponds to that with $L_c\sim 260$ and $A^\tau\sim 0.5$ at 
the signal dominant scale ($L\ll L_c$). 
For larger $R$, the amplitude becomes increased while $L_c$ is not significantly changed. 
Thus, $R\agt 10$ Mpc is not favored by the measurement of the Planck CMB temperature trispectrum. 
The typical bubble size at $x_e\sim 0.5$ indicated by simulations is 
$R\sim 1$-$10$ Mpc \cite{Furlanetto:2006,Zahn:2006b}. 
Consequently, the CMB trispectrum data starts to provide meaningful constraints on the patchy reionization. 

\begin{figure}[t]
\bc
\includegraphics[width=8.9cm,height=6.5cm,clip]{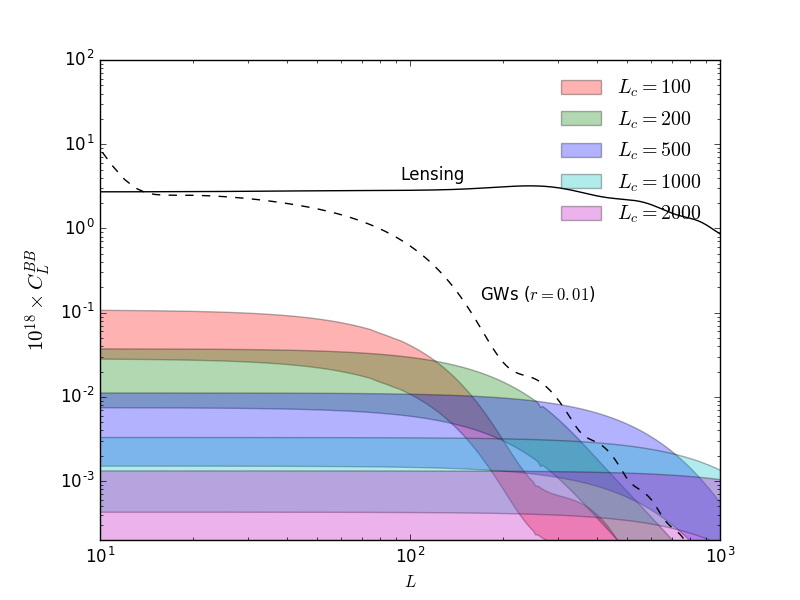} 
\caption{
The $B$-mode spectrum from the patchy reionization assuming the model of Ref.~\cite{Dvorkin:2009ah} 
and the constraints on the optical-depth power spectrum obtained in this work. 
We also show the $B$-mode spectra from the lensing and primordial gravitational waves 
with the tensor-to-scalar ratio, $r=0.01$. 
}
\label{fig:bb}
\ec
\end{figure}

The CMB $B$-mode power spectrum could be also generated by the patchy reionization 
and some previous works claim that such $B$ modes could dominate over the lensing $B$ modes 
\cite{Dore:2007,Dvorkin:2009ah,Mortonson:2010}. 
This also indicates that the patchy-reionization $B$-modes could be a possible contamination to 
detect the primordial gravitational waves (GWs) with the tensor-to-scalar ratio, $r\agt0.01$. 
Fig.~\ref{fig:bb} shows the $B$-mode spectrum from the patchy reionization, 
assuming the model of Ref.~\cite{Dvorkin:2009ah}. 
The power spectrum from the patchy reionization 
is roughly an order of magnitude smaller than the primordial GW $B$-mode spectrum with $r=0.01$ at $L=\mC{O}(10)$. 
The constraint above implies that large-scale $B$ modes from the patchy reionization are not 
a significant contamination to detect the primordial gravitational waves if $r\agt0.001$.

\section{Consistency and null tests} \label{discuss}

In the followings, we test potential residual systematics in our analysis:

\begin{table}
\centering
\caption{
Probability to exceed $\chi$ and $\chi^2$ statistic for alternative $C_L^{\tau\tau}$ measurements.
Note that the tests are not statistically independent and some of the tests are highly correlated. 
}
\label{table:PTE}
\begin{tabular}{lcc} \hline
 & $\chi$ & $\chi^2$ \\ \hline
\multicolumn{3}{c}{\it Baseline} \\
\SMICA & 0.81 & 0.15 \\
\multicolumn{3}{c}{\it Alternative CMB simulation} \\
\SMICA-FFP8 & 0.35 & 0.26 \\
\multicolumn{3}{c}{\it Foreground test} \\
Gal60 + Lensing mask   & 0.81 & 0.14 \\
\NILC-FFP8 & 0.20 & 0.29 \\
\multicolumn{3}{c}{\it Noise test} \\
$\tau^{\rm HM1}\times\tau^{\rm HM2}$ & 0.71 & 0.18 \\
\hline
\end{tabular}
\end{table}

\begin{table}
\centering
\caption{
Probability to exceed $\chi$ and $\chi^2$ statistics for the null tests.
}
\label{table:PTE:null}
\begin{tabular}{lcc} \hline
 & $\chi$ & $\chi^2$ \\ \hline
Reconstruction from HM half-differenced map & 0.20 & 0.81 \\
Difference of reconstructed $\tau$ ($(\tau^{\rm HM1}-\tau^{\rm HM2})/2$) & 0.37 & 0.25 \\ 
\hline
\end{tabular}
\end{table}

\bi
\item {\it Residual foreground tests} --- 
The results using the \SMICA\ foreground-cleaned map would be much less affected by foreground contaminations 
compared to that using a single frequency map. 
The residual foreground, however, may have an impact on our analysis. 
To test the residual foreground, we perform the same analysis but using \NILC\ data. 
We also change the mask region. 
The PTE values are within $2\sigma$. 
\item {\it Noise tests} --- 
We applied the noise scaling in the baseline analysis. 
We test this noise assumption by 1) measuring $\tau$ from the HM half-differenced map 
which contains only noise, and 2) computing cross correlation of $\tau$ reconstructed from 
the first and second halves of the HM data. 
Since the noise is almost uncorrelated between two data sets observed during different years, 
the cross correlation of $\tau$ is expected to be noise-independent. 
We find that the PTE values are consistent with null at $2\sigma$ level.
\item {\it Amplitude modulation} ---
If the amplitude of temperature anisotropies has a small uncertainty which varies randomly across the sky, 
e.g., gain fluctuations, the temperature fluctuations have an additional convolution term, 
$\hT(\hatn)=[1+\epsilon(\hatn)]\T(\hatn)$. These anisotropies mimic the effect of the patchy screening. 
To test the modulation of amplitude, we reconstruct $\tau$ from each HM data and differentiate them.
The random modulation leads to a nonvanishing signal in the difference, 
$(\tau^{HM1}-\tau^{HM2})/2$, while the true signal does not appear in the difference. 
The resultant PTE value is in agreement with the null hypothesis at $2\sigma$. 
\ei

\section{Summary and discussion} \label{summary}

We measured the Planck temperature trispectrum arising from the anisotropies in the optical depth. 
We showed that the observed trispectrum is consistent with that from the standard lensed CMB simulation 
within the $2\sigma$ level.
While no evidence from the patchy reionization was found in the Planck temperature trispectrum, 
the CMB constraint on the inhomogeneous reionization is significantly 
improved from the previous works. 
Assuming the model of Ref.~\cite{Wang:2006}, the constraints on the inhomogeneous reionization 
obtained in this work rule out a larger bubble size, $R\agt 10$ Mpc, at $2\sigma$ level. 
Further, our constraint implies that large-scale $B$-modes from the patchy reionization are not a significant 
contamination in detecting the primordial gravitational waves of $r\agt0.001$ 
if the $B$ mode induced by the patchy reionization is described by Ref.~\cite{Dvorkin:2009ah}.
We also examined potential residual systematics in the trispectrum analysis, showing that 
these systematics would be not significant in the trispectrum. 

We use the Planck data up to $\l=2048$ where the noise comes to dominate over the temperature signal. 
In the future, the use of the temperature at very small scales $\l\gg 3000$ will further improve 
the sensitivity to the patchy reionization. In such a case, however, other secondary effects 
such as kSZ and unresolved point sources should be taken into account in the reconstruction of 
the optical-depth anisotropies. 
Let us discuss a mitigation strategy of these secondary effects. 
The temperature anisotropies by the kSZ effect are described as $\T^{\rm kSZ}(\hatn)
\simeq \tau'v(\hatn)$ where $v(\hatn)$ is the line-of-sight peculiar velocity and $\tau'$ is 
the optical depth of an electron cloud \cite{SZ:1980}. 
These kSZ-induced temperature anisotropies lead to an additional mean-field bias in \eq{Eq:estt} and 
trispectrum in \eq{Eq:4pt}, but these contributions 
(and in general any point-source like contaminations) would be mitigated by the method described in 
Refs.~\cite{Namikawa:2012,P13:phi,Osborne:2013nna} since the kSZ contaminations are considered 
as roughly random Gaussian fields at scales relevant to the reconstruction. 
The fluctuations of $\tau'$ could also lead to additional contaminations but are expected to be 
smaller than the leading-order trispectrum. 
The method here can be further generalized to the case if the kSZ-induced temperature 
anisotropies are given as an integral of the electron clouds in the line-of-sight as similar to 
Ref.~\cite{Dvorkin:2008tf}. 
\ifdefined \kSZall
The temperature anisotropies by the kSZ effect are roughly described as $\T^{\rm kSZ}(\hatn)
\sim [\tau'+\delta\tau'(\hatn)]v(\hatn)$ 
where $v(\hatn)$ is the line-of-sight velocity, and $\tau'$ and $\delta\tau'(\hatn)$ are 
the homogeneous and spatially-dependent parts of the optical depth \cite{SZ:1980}. 
The term, $\tau' v(\hatn)$, leads to an additional mean-field bias in \eq{Eq:estt} and 
trispectrum in \eq{Eq:4pt}, but these contributions 
(and in general any point-source like contaminations) would be mitigated by the method described in 
Refs.~\cite{Namikawa:2012,P13:phi,Osborne:2013nna} since the kSZ contaminations are considered 
as roughly random Gaussian fields at scales relevant to the reconstruction. 
Similarly, the term, $\delta\tau'(\hatn) v(\hatn)$, can be also mitigated by 
constructing a quadratic estimator of $\delta\tau'(\hatn)$. 
\fi
We will leave a test of the methodology described here for our future work.

In the near future, the trispectrum constraints obtained in this paper will be improved by using data from 
wide-field ground-based CMB experiments with high angular resolution such as 
BICEP Array \cite{BICEPArray}, Simons Observatory \footnote{https://simonsobservatory.org/}, 
SPT-3G \cite{SPT3G}, and CMB-S4 \footnote{https://cmb-s4.org/}. 
In the above discussion, the kSZ-induced temperature signal is considered 
as a contamination of the optical-depth power spectrum measurement, but can also be used to detect 
the kSZ effect from reionization and will provide valuable implications 
for inhomogeneities of the reionization process \cite{Smith:2016lnt}.

\begin{acknowledgments}
We would like to thank Chao-Lin Kuo for support of this work and 
Cora Dvorkin, Vera Gluscevic and Mark Kamionkowski for helpful comments. 
This research used resources of the National Energy Research Scientific Computing Center, 
which is supported by the Office of Science of the U.S. Department of Energy under Contract No. DE-AC02-05CH11231.
\end{acknowledgments}

\appendix

\section{Verification of Pipeline} \label{verify}

\begin{figure*}[t]
\bc
\includegraphics[width=8.9cm,height=6.5cm,clip]{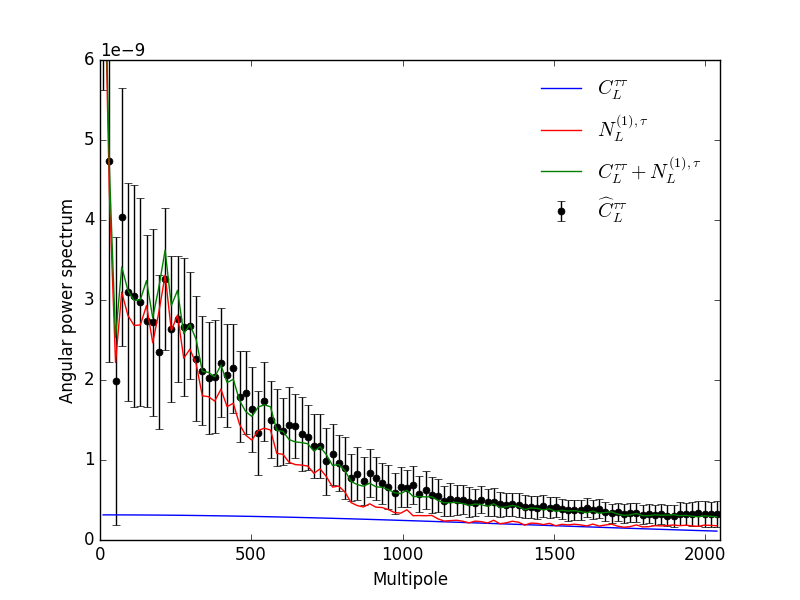} 
\includegraphics[width=8.9cm,height=6.5cm,clip]{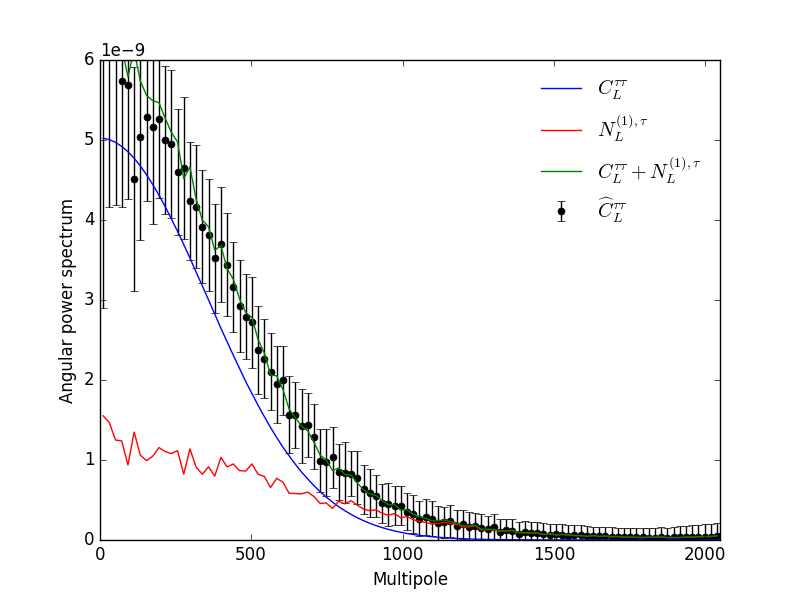} 
\caption{
{\it Left}: 
The power spectrum of the inhomogeneous reionization reconstructed from 
the simulated temperature map including the effect of the patchy reionization. 
We employ the lensed CMB signal and anisotropic noise maps of the \SMICA\ FFP8.1 simulation. 
The baseline mask (the lensing mask with the $1$deg apodization) is used. 
We assume $A^\tau=1$ and $L_c=2000$. 
{\it Right}: 
Same as Left but with $L_c=500$.
}
\label{fig:demo}
\ec
\end{figure*}

In this Appendix, we verify the pipeline used in this paper using the simulation including the effect 
of the inhomogeneous reionization. 

We first simulate the fluctuations of the optical depth, $\tau(\hatn)$, according to \eq{Eq:model} 
as a random Gaussian field. We use the \SMICA\ lensed CMB and noise simulation. 
The lensed CMB temperature map is modified by $1-\tau(\hatn)$. 
We reconstruct the inhomogeneous reionization from this modified simulation where the standard CMB temperature 
map is replaced with that including the effect of the inhomogeneous reionization. 

Fig.~\ref{fig:demo} shows the reconstructed power spectrum from the simulated temperature map. 
We also show the model spectrum of the inhomogeneous reionization given by \eq{Eq:model} and 
the secondary contraction of the temperature trispectrum, usually referred to as N1 bias, defined as 
\al{
	N_L^{(1),\tau} &= 
		\frac{1}{2}\ave{|\estt^{\T_{\bm 1}\T_{\bm 2}}_{LM}+\estt^{\T_{\bm 2}\T_{\bm 1}}_{LM}|^2}_{\bm 1,2,\tau}
	\notag \\
		&- \frac{1}{2}\ave{|\estt^{\T_{\bm 1}\T_{\bm 2}}_{LM}+\estt^{\T_{\bm 2}\T_{\bm 1}}_{LM}|^2}_{\bm 1,2}
	\,.
}
Here $\ave{\cdots}_{\bm 1,2,\tau}$ is the ensemble average over $i=\bm{1},\bm{2}$ with a fixed realization 
of $\tau$. 
The reconstructed spectrum from the simulation including the nonzero inhomogeneous $\tau$ 
corresponds to the sum of the input and N1 bias. 
The N1 bias in the reconstruction of the inhomogeneous reionization is significant compared to $C_L^{\tau\tau}$. 
We also perform the reconstruction with other values of $A^\tau$ and $L_c$, 
and find that the reconstructed spectrum is described as $C_L^{\tau\tau}+N_L^{(1,\tau)}$. 

\twocolumngrid
\bibliographystyle{apsrev}
\bibliography{exp,main}

\end{document}